\documentstyle[epsfig]{aipproc}

\newcommand{\pipipipi}{\mbox{$\pi^+\pi^-\pi^+\pi^-$ }}

\newcommand{\pipi}{\mbox{$\pi^{+}\pi^{-}$} }

\newcommand{\fmeson}{\mbox{$f_{2}(1270)$} }

\newcommand{\kkpi}{\mbox{$K^{0}_{S} K^{\pm} \pi^{\mp}$} }

\newcommand{\etapipi}{\mbox{$\mathrm{\eta \pi^{+} \pi^{-}}$} }

\begin{document}
\title{A search for non-$q \overline q$ mesons in the WA102 experiment at
the CERN Omega Spectrometer}

\author{Andrew Kirk$^*$ \\
and The WA102 collaboration$^{[1]}$}
\address{$^*$School of Physics and Astronomy, Birmingham University}

\maketitle

\begin{abstract}
A study of central meson production as a function of the
difference in transverse momentum ($dP_T$)
of the exchanged particles
shows that undisputed $q \overline q$ mesons
are suppressed at small $dP_T$ whereas the glueball candidates
are enhanced.
\end{abstract}
\begin{centerline}
{\it Invited talk at Hadron 97, August 1997}
\end{centerline}
\section*{Introduction}
\par
There is considerable current interest in trying to isolate the lightest
glueball.
Several experiments have been performed using glue-rich
production mechanisms.
One such mechanism is Double Pomeron Exchange (DPE) where the Pomeron
is thought to be a multi-gluonic object.
Consequently it has been
anticipated that production of
glueballs may be especially favoured in this process~[2].
\par
The Omega central production experiments
(WA76, WA91 and WA102) are
designed to study exclusive final states
formed in the reaction
\begin{center}
pp$\longrightarrow$p$_{f}X^{0}$p$_s$,
\end{center}
where the subscripts $f$ and $s$ refer to the fastest and slowest
particles in the laboratory frame respectively and $X^0$ represents
the central system. Such reactions are expected to
be mediated by double exchange processes
where both Pomeron and Reggeon exchange can occur.
\par
The trigger was designed to enhance double exchange
processes with respect to single exchange and elastic processes.
Details of the trigger conditions, the data
processing and event selection
have been given in previous publications~[3].
\section*{A Glueball-$q \overline q$ filter in central production ?}
\begin{figure}[b]
\begin{center}
\epsfig{file=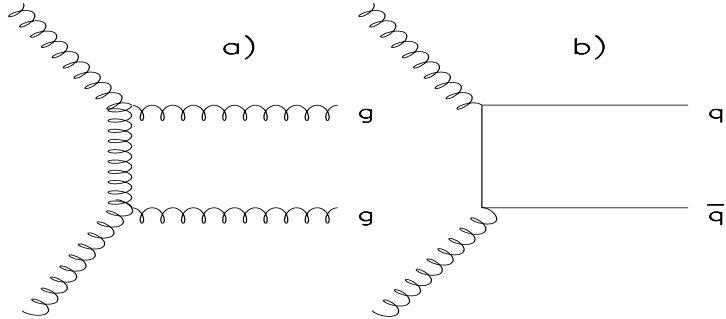,height=6cm,width=10.0cm,bbllx=0pt,bblly=0pt,
bburx=550pt,bbury=500pt}
\end{center}
\caption{Schematic diagrams
of the coupling of the exchange particles into the final state meson
for a) gluon exchange and b) quark exchange.}
\label{fi:feyn1}
\end{figure}
\par
The experiments have been
performed at incident beam momenta of 85, 300 and 450 GeV/c, corresponding to
centre-of-mass energies of
$\sqrt{s} = 12.7$, 23.8 and 28~GeV.
Theoretical
predictions~[4] of the evolution of
the different exchange mechanisms with centre
of mass energy, $\sqrt{s}$, suggest that
\begin{center}
$\sigma$(RR) $\sim s^{-1}$,\\
$\sigma$(RP) $\sim s^{-0.5}$,\\
$\sigma$(PP) $\sim$ constant,
\end{center}
where RR, RP and PP refer to Reggeon-Reggeon, Reggeon-Pomeron and
Pomeron-Pomeron
exchange respectively. Hence we expect Double Pomeron Exchange
(DPE) to be more significant at high energies, whereas the Reggeon-Reggeon and
Reggeon-Pomeron mechanisms will be of decreasing importance.
The decrease of the non-DPE cross section with energy can be inferred
from data
taken by the WA76 collaboration using pp interactions at $\sqrt{s}$ of 12.7 GeV
and 23.8 GeV~[5].
The \pipi mass spectra for the two cases show that
the signal-to-background ratio for the $\rho^0$(770)
is much lower at high energy, and the WA76 collaboration report
that the ratio of the $\rho^0$(770) cross sections at 23.8 GeV and 12.7 GeV
is 0.44~$\pm$~0.07.
Since isospin 1 states such as the $\rho^0$(770) cannot be produced by DPE,
the decrease
of the $\rho^{0}(770)$ signal at high $\sqrt{s}$
is consistent with DPE becoming
relatively more important with increasing energy with respect to other
exchange processes.
\begin{figure}[b]
\begin{center}
\epsfig{file=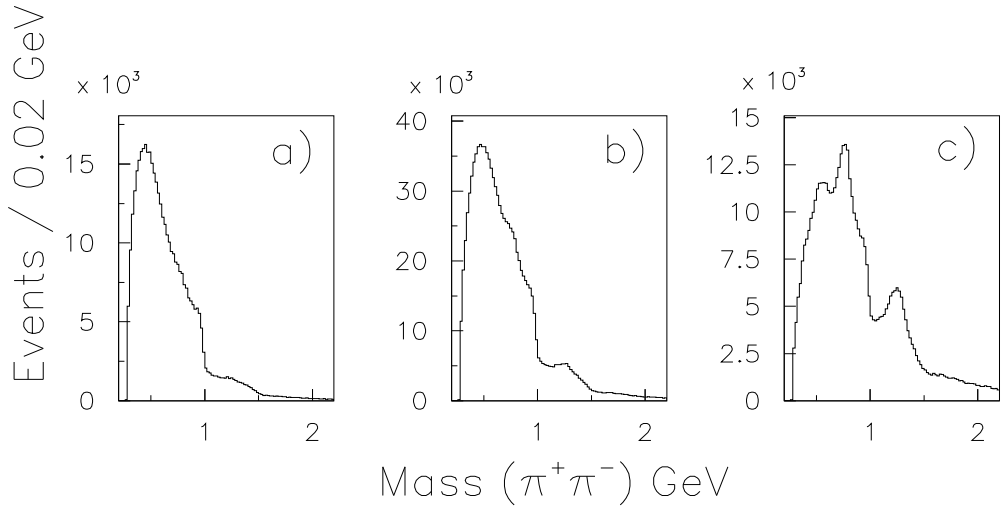,height=8cm,width=14.0cm,bbllx=0pt,
bblly=0pt,bburx=680pt,bbury=400pt}
\end{center}
\caption{ The \pipi mass spectrum for
a) $dP_T <   0.2$ GeV, b) $0.2 < dP_T <   0.5$ GeV and c) $dP_T >   0.5$ GeV.}
\label{fi:2pi}
\end{figure}
\begin{figure}[b!]
\begin{center}
\epsfig{file=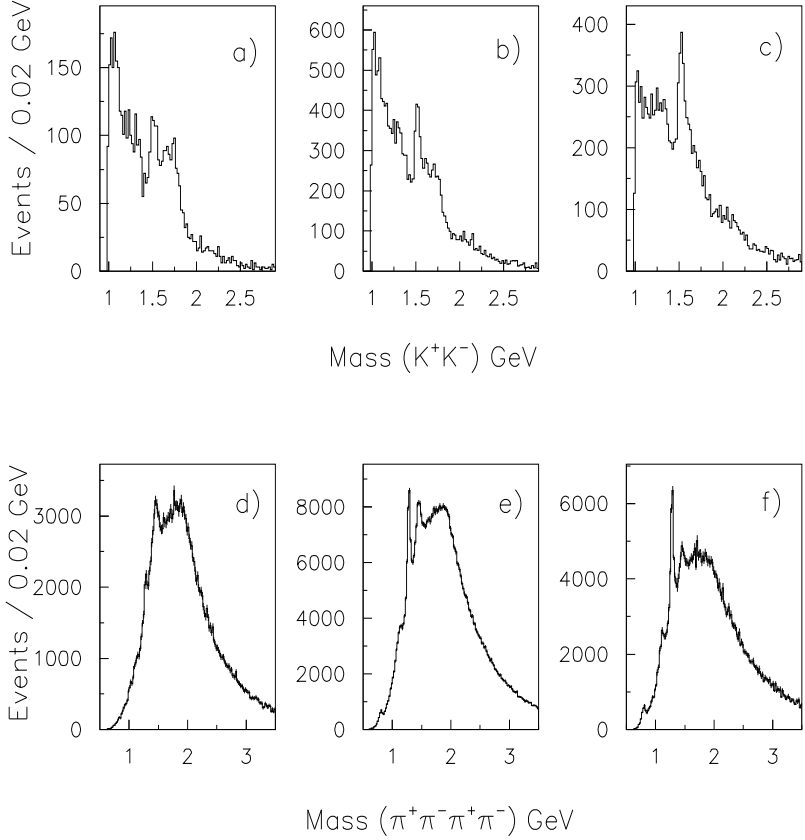,height=14cm,width=14cm}
\end{center}
\caption{$K^+K^-$ mass spectrum for a) $dP_T <   0.2$ GeV,
b) $0.2 <   dP_T <   0.5$ GeV and c) $dP_T >   0.5$ GeV and
the \pipipipi mass spectrum for d) $dP_T <   0.2$ GeV,
e) $0.2 <   dP_T <   0.5$ GeV and f) $dP_T >   0.5$ GeV.}
\label{fi:2k}
\end{figure}
\par
However,
even in the case of pure DPE
the exchanged particles still have to couple to a final state meson.
The coupling of the two exchanged particles can either be by gluon exchange
or quark exchange. Assuming the Pomeron
is a colour singlet gluonic system if
a gluon is exchanged then a gluonic state is produced, whereas if a
quark is exchanged then a $q \overline q $ state is produced
(see figures~\ref{fi:feyn1}a) and b) respectively).
It has been suggested recently~[6] that
for small differences in transverse momentum between the two
exchanged particles
an enhancement in the production of glueballs
relative to $q \overline q$ states may occur.
\par
Recently the WA91 collaboration has published a paper~[7]
showing that the observed centrally produced resonances depend on the
angle between the outgoing slow and fast protons.
In order to describe the data in terms of a physical model,
Close and Kirk~[6],
have proposed that the data be analysed
in terms of the difference in transverse momentum
between the particles exchanged from the
fast and slow vertices.
The difference in the transverse momentum vectors ($dP_T$) is defined to be
\begin{center}
$dP_T$ = $\sqrt{(P_{y1} - P_{y2})^2 + (P_{z1} - P_{z2})^2}$
\end{center}
where
$Py_i$, $Pz_i$ are the y and z components of the momentum
of the $ith$ exchanged particle in the pp centre of mass system~[8].
\par
The effect
that different cuts in $dP_T$ have on the $\pi^+\pi^-$ mass spectrum are shown
in
figures~\ref{fi:2pi}a), b) and c).
As can be seen, for $dP_T$~$<$~0.2~GeV there is effectively
no $\rho^0$(770) or \fmeson signals. These signals only become
apparent as $dP_T$ increases.
However the $f_0(980)$, which is responsible for the sharp drop in the
spectrum around 1~GeV, is clearly visible in the small
$dP_T$ sample.
\par
Figures~\ref{fi:2k}a), b) and c) show the effect of the
$dP_T$ cut on the $K^+ K^-$ mass spectrum where
structures can be observed in the 1.5 and 1.7 GeV mass region which have
been previously identified as the
$f_{2}^\prime$(1525) and the $f_J(1710)$~[9].
As can be seen,
the $f_{2}^\prime$(1525) is produced dominantly at high $dP_T$,
whereas the $f_J(1710)$ is produced dominantly at low $dP_T$.
\par
In the \pipipipi  mass spectrum a dramatic effect is observed,
see figures~\ref{fi:2k}d), e) and f).
The $f_1(1285)$ signal has virtually disappeared at low $dP_T$
whereas
the $f_0(1500)$ and $f_2(1930)$ signals remain.
\begin{figure}[b!]
\begin{center}
\epsfig{file=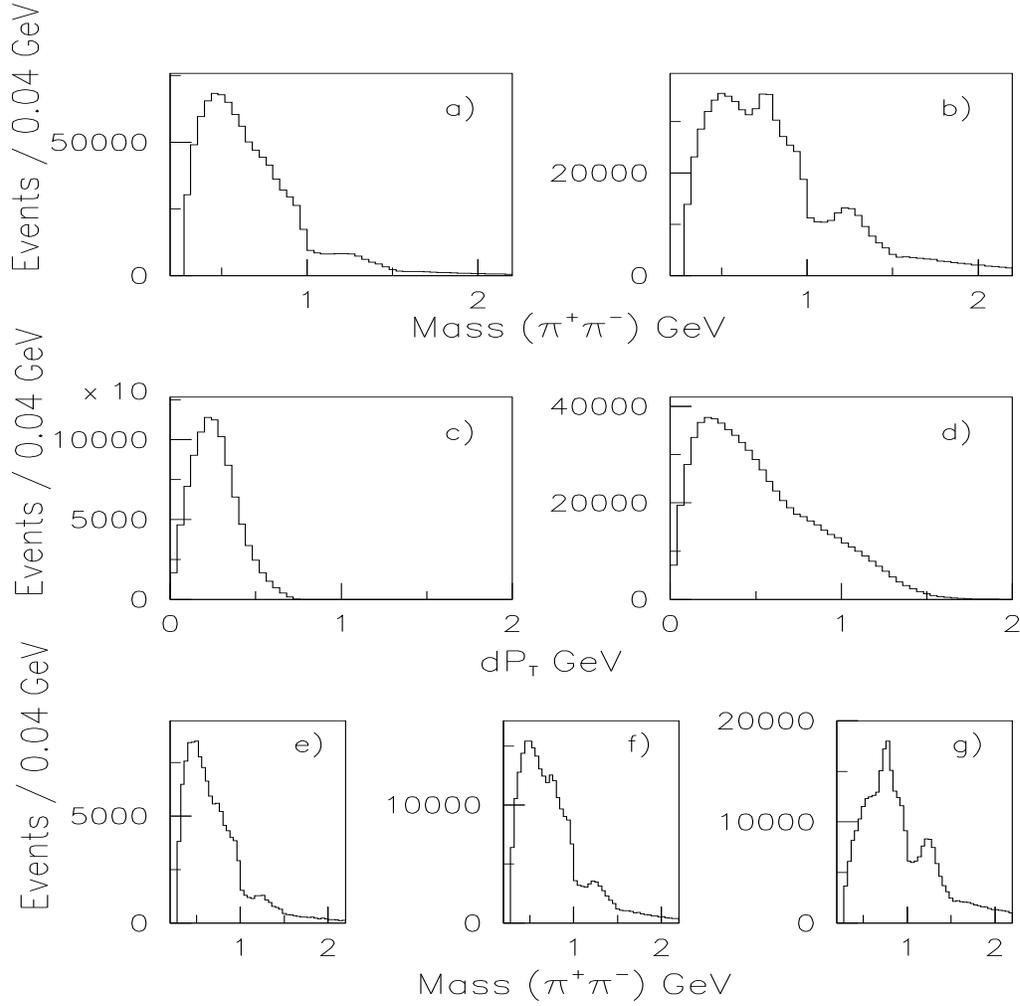,height=14cm,width=14cm}
\end{center}
\caption[tcut]{Results of cutting on the four momentum transferred at the
proton
vertices.
The $\pi^+\pi^-$ mass spectrum for a) $|t_f|$ $<$ 0.15 and $|t_s|$ $<$ 0.15
GeV$^2$ and
c) $|t_f|$ $>$ 0.15 and $|t_s|$ $>$ 0.15 GeV$^2$.
The $dP_T$ distribution for c) $|t_f|$ $<$ 0.15 and $|t_s|$ $<$ 0.15 GeV$^2$
and
d) $|t_f|$ $>$ 0.15 and $|t_s|$ $>$ 0.15 GeV$^2$.
The $\pi^+\pi^-$ mass spectrum for $|t_f|$ $>$ 0.15 and $|t_s|$ $>$ 0.15
GeV$^2$ and
e) $dP_T$ $<$ 0.2 GeV, f) 0.2 $<$ $dP_T$ $<$ 0.5 GeV and g) $dP_T$ $>$ 0.5 GeV.
}
\label{fi:tdep}
\end{figure}
\begin{figure}
\begin{center}
\epsfig{file=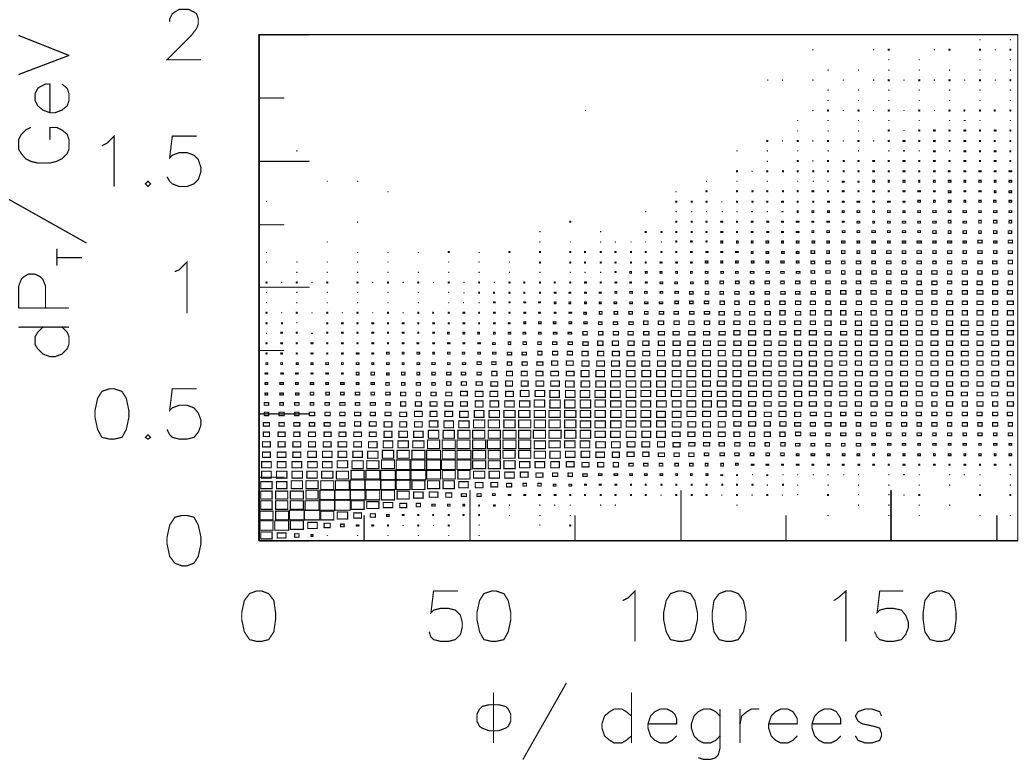,height=10cm,width=14cm}
\end{center}
\caption{$dP_T$ versus the azimuthal angle between the fast and
slow protons ($\phi$).
}
\label{fi:phidep}
\end{figure}
\par
A spin-parity analysis of the
$\pi^{+}\pi^{-}\pi^{+}\pi^{-}$
channel
has been performed~[10] using an isobar model~[11].
The $f_{1}(1285) $ is clearly seen in the
$J^{P}=1^{+}$~$\rho\rho $
and the $f_1(1285)$ signal almost disappears at
small $dP_T$.
In the $J^{P} =0^{+} \rho\rho $ distribution
a peak is observed at 1.45~GeV
together with a broad enhancement around 2~GeV.
The peak
in the $J^{P} =0^{+} \rho\rho $ wave
around 1.45 GeV remains
for $dP_T$~$\leq$~0.2~GeV while the
$J^{P} =0^{+} $
enhancement at 2.0~GeV becomes less
important: which shows that the $dP_T$ effect is not simply a $J^P$ filter.
\par
A fit has first been performed to the total
$J^{P} =0^{+} \rho\rho $
distribution using a K matrix formalism~[12]
including poles to describe the peak at 1.45 GeV as an interference between the
$f_0(1300)$, the $f_0(1500)$ together with a possible state at 2~GeV.
The resulting resonance parameters for
the $f_0(1300)$ and
$f_0(1500)$ are very similar to those found by Crystal Barrel~[13].
\par
The peak observed at 1.9~GeV, called the $f_2(1900)$,
is found to
decay to $a_{2}(1320)\pi$
and $f_{2}(1270)\pi\pi$ with $J^{PC}=2^{++}$.
At small $dP_T$ the $f_2(1900)$ signal is still important.
This is the first evidence of a non-zero spin resonance produced
at small $dP_T$ and hence shows that the $dP_T$ effect is not just
a $J^P$~=~$0^+$ filter.
\par
In addition to these waves,
a $J^{P} =2^{-}$ $a_{2}(1320)\pi $ wave was required in the fit.
The
$J^{P} =2^{-}$ $a_{2}(1320)\pi $ wave observed in this experiment is consistent
with the two $\eta_2$ resonances observed by Crystal Barrel~[14]
with both states decaying
to $a_2(1320) \pi $.
The $2^{-+} a_2(1320)\pi$ signal is
suppressed at small $dP_T$. This behaviour is
consistent with the signals being due to
standard $q \overline q$ states~[6].
\par
A similar effect is observed in the \kkpi~[15]
and \etapipi~[8] channels
where the
$f_1(1285)$,
$f_1(1420)$ and $\eta^{\prime}$ are all more prominent in the
large $dP_T$ sample and start to
disappear at low $dP_T$.
\par
In fact it has been observed that
all the undisputed
$ q \overline q $ states
(i.e. $\rho^0(770)$, $\eta^{\prime}$, \fmeson, $f_1(1285)$,
$f_2^\prime(1525)$ etc.)
are suppressed as $dP_T$ goes to zero,
whereas the glueball candidates
$f_J(1710)$, $f_0(1500)$ and $f_2(1930)$ survive.
It is also interesting to note that the
enigmatic
$f_0(980)$,
a possible non-$q \overline q$ meson or $K \overline K$ molecule state does not
behave as a normal $q \overline q$ state.
\begin{figure}[hbt]
\begin{center}
\epsfig{file=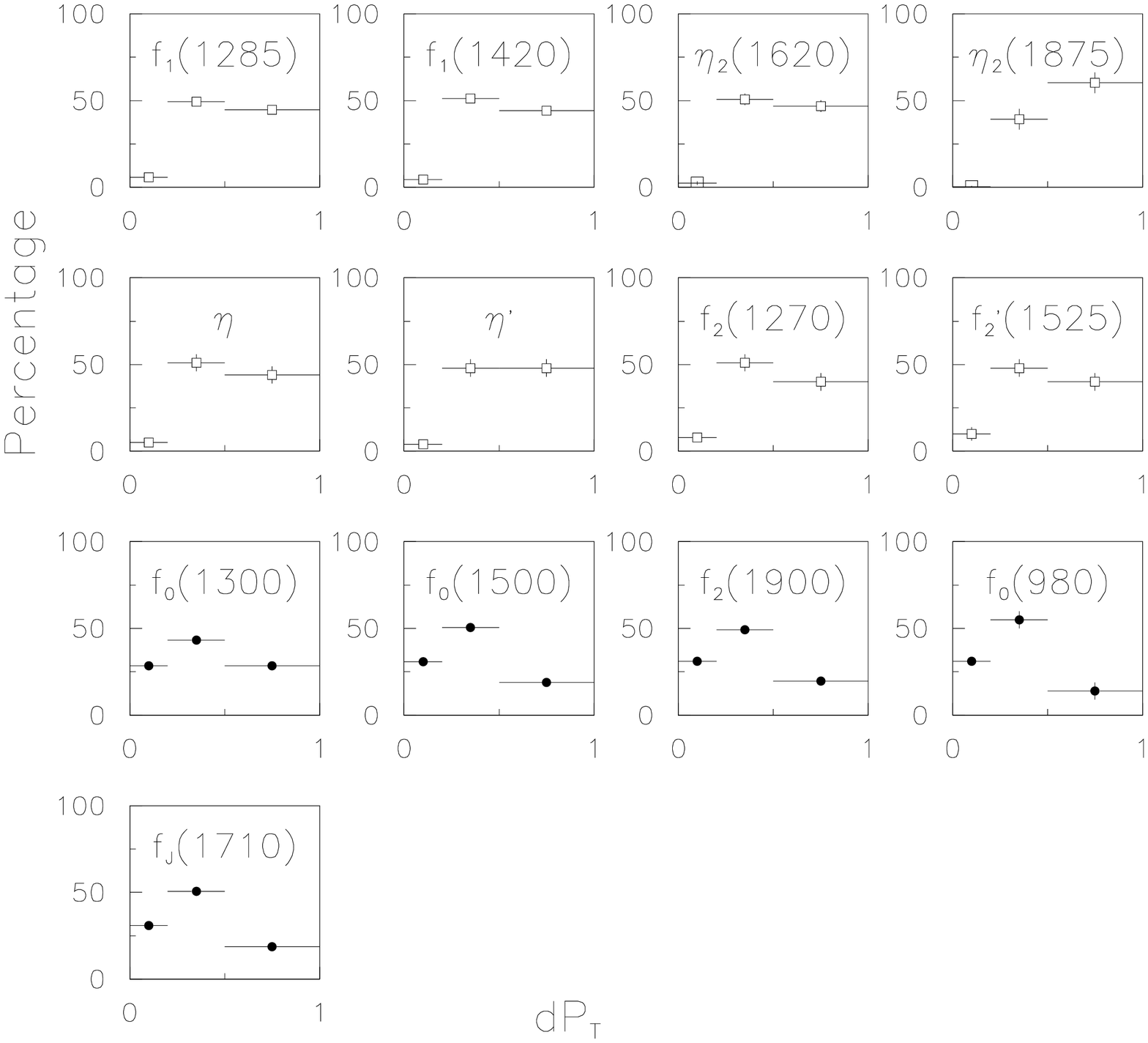,height=14cm,width=14cm}
\end{center}
\caption{The percentage of the resonance as a function of $dP_T$}
\label{frac}
\end{figure}
\begin{figure}[t!]
\begin{center}
\epsfig{file=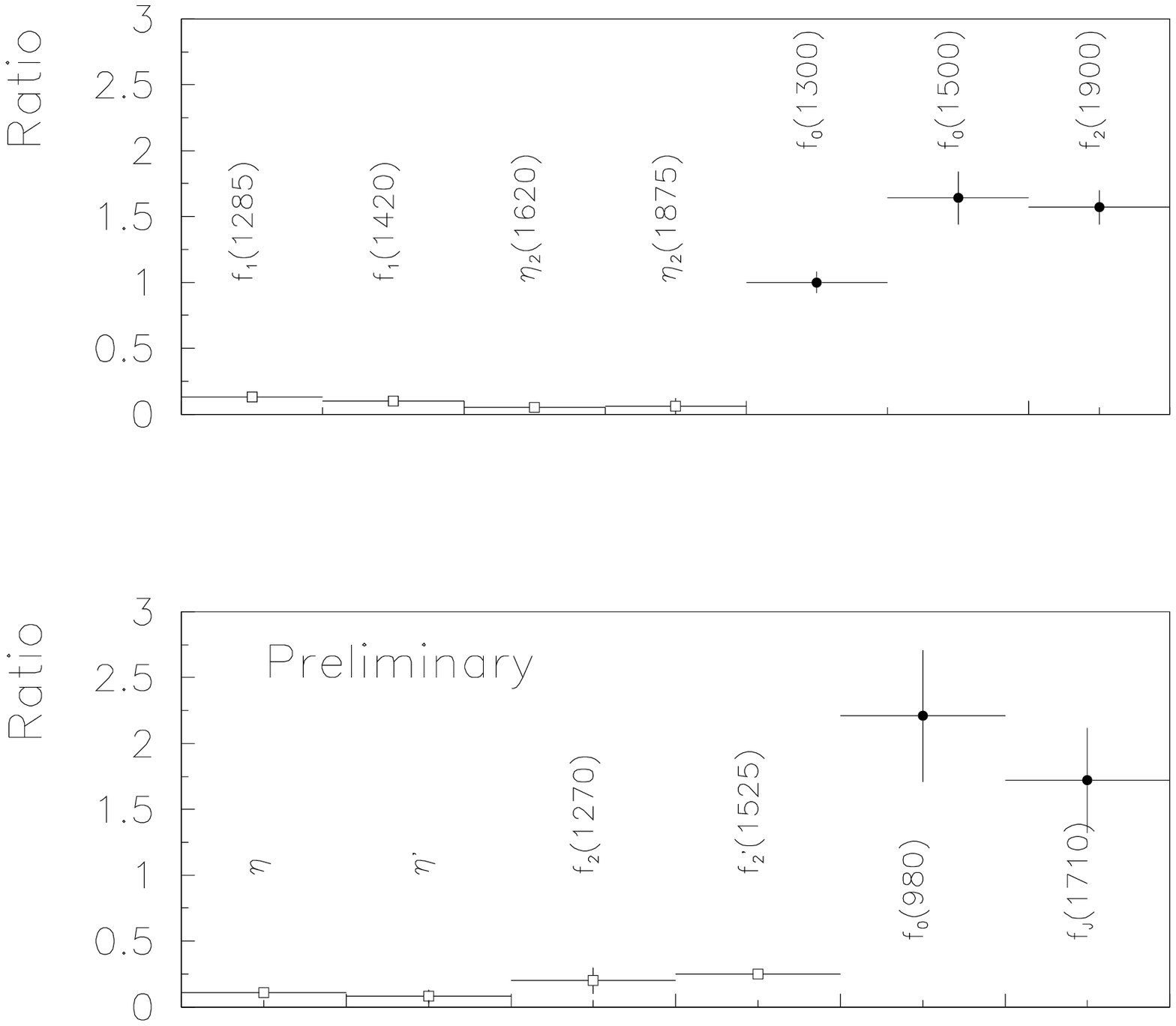,height=14cm,width=14cm}
\end{center}
\caption{The ratio of the amount of resonance with
$dP_T$~$\leq$~0.2 to the amount with
$dP_T$~$\geq$~0.5~GeV.
}
\label{fracratio}
\end{figure}
\par
A Monte Carlo simulation of the trigger, detector acceptances
and reconstruction program
shows that there is very little difference in the acceptance as a function of
$dP_T$ in the different mass intervals considered
within a given channel and hence the
observed differences in resonance production can not be explained
as acceptance effects.
\par
It has previously been observed that the resonances produced in the central
region depend on the four momentum transferred from the fast ($t_f$) and slow
vertices ($t_s$)~[5].
The $\pi^+\pi^-$ mass spectrum is shown for the case where $|t_f|$ and $|t_s|$
are both less than 0.15 GeV$^2$ in figure~\ref{fi:tdep}a) and in
figure~\ref{fi:tdep}b) for the case when
$|t_f|$ and $|t_s|$ are both greater than 0.15 GeV$^2$. As can be seen
the amount of $\rho^0(770)$ and $f_2(1270)$
does change as a function of this cut.
However,
in figures~\ref{fi:tdep}c) and d) the $dP_T$ distribution for these two cases
is shown where it can be seen the events that have small $|t|$ are restricted
to small values of $dP_T$. To show that $dP_T$ is the
most important underlying dynamical effect the
$dP_T$ cut has been applied to the sample of events with large $|t|$.
Figures~\ref{fi:tdep}e), f) and g) show the events when
$|t_f|$ and $|t_s|$ are both greater than 0.15 GeV$^2$ for
$dP_T$~$\leq$~0.2~GeV,
0.2~$\leq$~$dP_T$~$\leq$~0.5~GeV and
$dP_T$~$\geq$~0.5~GeV respectively.
As can be seen the $dP_T$ cut still works in this sample and hence it
would seem that $dP_T$ is the most important cut to be used.
\par
An effect on the resonances observed has also been seen when cuts have been
made on
the azimuthal angle ($\phi$) between the fast and slow proton.
This angle $\phi$ is related to $dP_T$ by
\begin{center}
$cos \phi  = \frac{ dP_T^2  -  P_T^2}{4 t_s t_f}$
\end{center}
where $P_T$ is the transverse momentum of the central system.
The correlation between $dP_T$ and $\phi$ is shown in figure~\ref{fi:phidep}.
Although cuts in $\phi$ do produce an effect on the resonances
observed the effect is not as clear compared to cuts in $dP_T$.
\section*{Summary of the effects of the $dP_T$ filter}
\par
In order to calculate the contribution of each resonance as a function
of the $dP_T$ the mass spectra have been fitted with
the parameters of the resonances fixed to those obtained from the
fits to the total data. The results of these fits
are summarised in
figure~\ref{frac} where the percentage of each resonance as a function of
$dP_T$ is presented.
Figure~\ref{fracratio} shows the ratio of the number of events
for $dP_T$ $<$ 0.2 GeV to
the number of events
for $dP_T$ $>$ 0.5 GeV for each resonance considered.
As can be seen all the undisputed
$q \overline q$ states have a small value for this ratio whereas
the interesting states have a high value.

\section*{Conclusions}
\par
Preliminary results show that there is the possibility of a
glueball-$q \overline q$ filter mechanism in central production.
All the
undisputed $q \overline q $ states are observed to be suppressed
at small $dP_T$, but the glueball candidates
$f_0(1500)$, $f_J(1710)$, and $f_2(1930)$ ,
together with the enigmatic $f_0(980)$,
survive.


%

\end{document}